\documentclass[12pt]{article}
\usepackage[utf8]{inputenc}
\usepackage{hyperref}
\usepackage[margin=1in]{geometry}
\usepackage{setspace}
\usepackage{color}
\usepackage{braket}
\usepackage{titling}
\usepackage{graphicx}
\usepackage[bf,margin=20pt,font={small}]{caption}
\usepackage{subcaption}
\usepackage{tikz}
\usepackage{youngtab}
\usepackage{epigraph}
\usetikzlibrary{math}
\newcommand{\eps}{\epsilon}
\usepackage{amsmath,amssymb,dsfont}
\allowdisplaybreaks
\newcommand{\tr}{\operatorname{tr}}
\newcommand{\Tr}{\operatorname{tr}}
\newcommand{\ad}{\operatorname{ad}}
\newcommand{\gtwo}{g_{\rm YM}^2}
\newcommand{\D}{\mathcal{D}}
\newcommand{\itd}{\int d^2 x}
\usepackage{tikz}

\definecolor{royalblue}{rgb}{0.00000,0.44700,0.74100}%
\definecolor{royalorange}{rgb}{0.85000,0.32500,0.09800}%
\definecolor{royalyellow}{rgb}{0.92900,0.69400,0.12500}%
\definecolor{purple}{rgb}{0.5804, 0.0, 0.82745098}%
\definecolor{applegreen}{rgb}{0.55, 0.71, 0.0}
\definecolor{bittersweet}{rgb}{1.0, 0.44, 0.37}

\usetikzlibrary{external}
\tikzsetexternalprefix{images/}
\tikzset{external/system call={lualatex -shell-escape -halt-on-error -interaction=batchmode -jobname "\image" "\texsource"}}
\DeclareMathAlphabet{\mathpzc}{OT1}{pzc}{m}{it}

\usepackage{putex}

\usepackage{mathtools}
\usepackage{siunitx}

\usepackage{pgfplots}
\pgfplotsset{compat=1.10}
\usepgfplotslibrary{fillbetween}
\usetikzlibrary{patterns}

\institution{NYU}{CCPP, Department of Physics, NYU, New York, NY,  10003, USA}

\title{Supersymmetry in QCD$_2$ coupled to fermions}

\authors{Fedor K.~Popov\worksat{\NYU}}

\abstract{We consider $1+1$ dimensional Yang-Mills theory with gauge group $G$ coupled to a massive Majorana fermion field in an adjoint representation and a number of massless Dirac or Majorana fermions transforming in arbitrary representations of the gauge group $G$. It is shown that the spectrum of the massive sector of this theory becomes supersymmetric at particular mass of adjoint fermion. This mass is independent of the detailed structure of the massless sector of the model and depends only on the gauge group $G$ and integer $k$ measuring the total anomaly. The massless sector of the model is shown to be not necessarily supersymmetric.
}

\date{\today}

\begin{document}
\maketitle
\tableofcontents
\section{Introduction and Supersymmetric Quantum Electrodynamics}

One of the most important enigmas of the modern theoretical physics is the problem of confinement in Quantum Chromodynamics (QCD), that is just a $SU(3)$ Yang-Mills theory coupled to fermions in fundamental representation. Due to experimental advances we know that these fermions are bound into hadrons and never observed as single particles. Even though we understand the rules and laws governing the behavour of quarks at small distances, we still does not completely comprehend the QCD and its properties at large distances.  One of the promising ideas that allows to drastically simplify the problem and give some understanding was proposed by G.'t-Hooft \cite{t1993planar}. Namely, he suggested an idea, that if we consider $SU(N)$ gauge theory at large $N$ limit with fixed $\lambda=g_{\rm YM}^2 N$, then only some particular type of diagrams would contribute. To demonstrate the power of such an approach 't-Hooft managed to solve an $SU(N)$ gauge theory in $1+1$ dimensions  that is coupled to  a massive Dirac fermion field in the fundamental representation \cite{t1993two}. The various generalizations of 't-Hooft model were considered, for instance, coupling to fermion fields in different representations of $SU(N)$. In the large $N$ limit these models could not be solved in a similar fashion, but still some other interesting and peculiar models were proposed and studied.

In this short letter we discuss the properties of spectrum of these two-dimensional models of quantum chromodynamics. Namely, we will be interested in the question, when such models could become supersymmetric. For this purpose, we consider a gauge field with gauge group $G$, that is assumed to be a simple Lie group, and couple it to a massive Majorana fermion field $\psi^i$ that belongs to an adjoint representation of the gauge group $G$. One of the remarkable properties that the spectrum of the model becomes supersymmetric at particular mass of the fermions $m_{\rm adj}$. It was shown first  by Kutasov \cite{kutasov1994two} and after that checked explicitly using some numerical methods \cite{Bhanot:1993xp,Dempsey:2021xpf,Gross:1997mx}. Even though this property of two dimensional gauge theories was very well established, the physical intuition behind this result is still obscure and understood only in terms of light-cone quantization. We propose a claim, that this supersymmetric QCD models could be understood as the deformation of $\mathcal{N}=1$ supersymmetric WZW models by relevant operators. Namely, let us consider a WZW model or some representation of Kac-Moody algebra at level $\tilde{k}$:
\begin{gather}
	\left[\tilde{J}^a_n, \tilde{J}^b_m\right] = i f^{abc} \tilde{J}^c_{n+m} +  n\delta_{n,-m} \delta_{ab} \frac{\tilde{k}}{2}.
\end{gather}
It is easy to make the model supersymmetric by introducing Majorana fermions $\psi^i$ in the adjoint representation \cite{kazama1989characterization,Witten:1991mk}. Then we should redefine the operators $\tilde{J}^a_n$ so they would correctly act on fermion operators $\psi^i_n$ and construct a fermionic operator $Q$:
\begin{gather}
	\left\{\psi^i_n,\psi^j_m \right\} = \delta^{ij} \delta_{n+m,0}, \quad J^a_n = \tilde{J}^a_n + \frac{i}{2} \sum_m f_{abc}\psi^b_{n-m}\psi^c_m,\quad \left[ J^a_n, \psi^b_m\right] = if^{abc}\psi^c_{n+m} \notag\\
	 Q=\sum_{n,m}\frac{1}{6} f_{abc} \psi^a_n \psi^b_m\psi^c_{-n-m} + \sum_n J_{n}^i \psi^i_{-n}, \label{eq:WZWsuper}
\end{gather}
it is easy to check that $Q$ would interchange the current and fermionic operators $J^i_n$
$$
	\left\{Q,\psi^i_n\right\} = J^i_n,\quad \left[Q,J^i_n\right] = i\frac{k n}{4\pi} \psi^i_n.
$$
Since $Q$ is a fermionic operator we conclude that $Q$ is a supersymmetry operator. One of the interesting features of this construction, that it does not depend on the initial level $k$ of the Kac-Moody algebra. We can just start with an empty representation of the Kac-Moody algebra ($k=0$) and just by considering fermions alone in an adjoint representation we can see that they automatically are supersymmetric.

The simplest example of such supersymmetry is just to consider a $U(1)$ version of Kac-Moody algebra and couple to an adjoint Majorana fermion, or just a free massless Majorana fermion. The $U(1)$ Kac-Moody algebra is just a Heisenberg-Weyl algebra and could be realized with a free scalar field. Therefore the supersymmetric $U(1)$ Kac-Moody algebra could be realized in the following way
\begin{gather}
S = \int d^2 x\left[\frac12 \left(\partial_\mu \phi\right)^2 + i \bar{\psi} \gamma^\mu \partial_\mu \psi \right],
\end{gather}
that is easily seen to be supersymmetric by construction (spectrum consists of a massless fermion and scalar field, therefore it is supersymmetric). We can make this model a little bit more complicating by introducing a mass term to both fields (that could be considered as a deformation of the initial CFT by an relevant operator, but that still respects a supersymmetry)
\begin{gather}
S = \int d^2 x\left[\frac12 \left(\partial_\mu \phi\right)^2 - \frac12 m^2 \phi^2 + i \bar{\psi} \gamma^\mu \partial_\mu \psi + i m \bar{\psi}\psi \right], \label{eq:massmodel}
\end{gather}
we can note that a massive scalar field could be realized as a $U(1)$ gauge field coupled to a massless fermion. Therefore we must conclude that the following model must be supersymmetric
\begin{gather}
S = \int d^2 x\left[-\frac{1}{4 e^2} F_{\mu\nu}^2 + i \bar{\chi}\gamma^\mu\left(\partial_\mu - i e A_\mu\right)\chi  + i \bar{\psi} \gamma^\mu \partial_\mu \psi + i \frac{e}{\sqrt{2\pi}} \bar{\psi}\psi \right],
\end{gather}
let us note that this particular model could serve as a simple example of the supersymmetric QCD model with massive fermions in the adjoint representation discovered by Kutasov \cite{kutasov1994two}. The main difference is that along with the fermions in adjoint representation we have added massless charged fermions, that essentially created mass for the scalar field in the action \eqref{eq:massmodel}. The consideration $U(1)$ gauge group is quite simple and tractable, that allows to understand the physics and mathematics behind this anomalous supersymmetry. Thus, while the classical supersymmetry requires the mass of a fermion field in the adjoint representation to be zero, the supersymmetric transformation are chiral and therefore measure is not invariant under the action of supersymmetry. To compensate this additional term we should add a mass term for a Majorana fermion. In the case of non-abelian gauge fields we could expect that the same reasoning happens but the computation becomes more complicated. Nonetheless, the essential peculiarities and properties are left the same. Comparison with $U(1)$ model allows us to make a quite interesting generalization of this supersymmetry, when along with massive fermions in adjoint representation we add some massless fermions in various representations of the gauge group and one can still find mass $m_{\rm adj}$, where the model becomes supersymmetric.

\section{The Hamiltonian and Hilbert Space of two dimensional QCD models}
As it was discussed in the introduction we want to study a gauge field theory could to a massive Majorana fermion in the adjoint representation of some continuous Lie group $G$ together with some massless Dirac or Majorana fermions in a different reducible unitary representation $\mathcal{R}$. For conventions and notations we refer to the appendix A. The action of the model is
\begin{gather}
	S = \int d^2 x\left(\tr\left[-\frac{1}{4 \gtwo} F_{\mu\nu}^2 + i \bar{\Psi} \gamma^\mu \D_\mu \Psi - m_{\rm adj}\bar{\Psi} \Psi \right] + i \bar{v}\gamma^\mu \D_\mu v\right) \label{eq:2dqcdaction}
\end{gather}
To make all computations tractable and explicit we will write it down directly in terms of the component of the fields
\begin{gather}
	S = \int d^2 x\left[\frac{1}{4 \gtwo} F^{i\, 2}_{+-}  + \left(\frac{i}{2} f_{ijk} \psi_-^i \psi_-^j +   \bar{v}_-^\alpha \tau^k_{\alpha\beta} v_-^\beta\right)A_+^k -\left(-\frac{i}{2}f_{ijk} \psi_+^i \psi_+^j + \bar{v}_+^\alpha \tau^k_{\alpha\beta}v_+^\beta  \right) A_-^k + \right.\notag\\\left.
	+ \frac{i}{2} \psi^i_+ \partial_- \psi^i_+ + \frac{i}{2}\psi^i_- \partial_+ \psi^i_+ + \frac{i}{\sqrt{2}} m_{\rm adj} \psi^i_+ \psi^i_- + i\bar{v}^\alpha_- \partial_+ v^\alpha_-  + \bar{v}^\alpha_+\partial_- v^\alpha_+ \right],
\end{gather}
that coincide with the action derived in \cite{Dempsey:2021xpf}. It is convenient now to study this model in the light-cone gauge where we treat one of the light-cone coordinates, for instance, $x_-$ as time and the other, $x^+$, as a spatial coordinate. In this case we pick the gauge $A^i_+ = 0$, that would make the other component field $A^i_-$ non-dynamical. Then its impact on the action could be easily taken into account.
The same holds for the fermionic fields $\psi^i_-$ and $v^\alpha_-$. After integrating out $\psi^i_-$ and $A_-$ we get the following action
\begin{gather}
S= \int d^2x \left[i \bar{v}^\alpha_+\partial_- v^\alpha_+ + \frac{i}{2} \psi^i_+ \partial_- \psi^i_+ - \gtwo J^i \frac{1}{\partial_+^2} J^i  - \frac{i m^2_{\rm adj}}{4} \psi^i \frac{1}{\partial_+} \psi^i   \right],\notag\\
 \text{where} \quad J^i =- \frac{i}{2} f_{ijk} \psi^j_+ \psi^k_+ + \bar{v}^\alpha_+ \tau^i_{\alpha\beta} v^\beta_+
\end{gather}
From this action we immediately see that the only dynamical fields are $\psi^i_+$ and $v^\alpha_+$. Then following the standard quantization procedure we construct the Hilbert space $\mathcal{H}$ and Hamiltonian that acts in this space.  Since the action contains only first derivative of the fermionic fields we have quite simple commutation relations
\begin{gather}\pi_\psi^i = \frac{\partial \mathcal{L}}{\partial\left(\partial_- \psi^i_+\right)} = i \partial_- \psi^i_+,\quad
\left.\left\{ \psi^i_+(x^+),\psi^j_+(y^+)\right\}\right. = \delta^{ij} \delta(x^+-y^+)\notag\\
\pi_v^\alpha = \bar{v}^\alpha,\quad \left.\left\{ \bar{v}^\alpha_+(x^+),v^\beta_+(y^+)\right\}\right. = \delta^{\alpha\beta} \delta(x^+-y^+) \label{eq:comrel}
\end{gather}
After that we can introduce the vacuum $\ket{0}$ as
\begin{gather}
	v^\alpha_+(k_+) \ket{0} = 0\quad \forall k_+ \in \mathbb{R}, \quad \psi_+^i(k_+) \ket{0}=0 \quad \forall k_+<0 \notag\\
	v^\alpha_+(k_+) = \int dx_+ v^\alpha_+(x_+) e^{- i k_+ x^+},\quad
	\psi^i_+(k_+) = \int dx_+ \psi^i_+(x_+) e^{- i k_+ x^+}
\end{gather}
and use the other fields $\bar{v}^\alpha_+(k_+)$ and $\psi^i_+(k_+)$ to act on the vacuum and $\ket{0}$ and construct the other states in the Hilbert space $\mathcal{H}$ in a way analogous to the Clifford module, where the finite number of Majorana or Dirac fermions is considered.

To make the statement more concrete we should take into account residual gauge symmetries. Namely, a light-cone gauge $A_+=0$ does not fix the gauge completely, we can still make a gauge transformation with parameter being a function of $x_-$ alone
\begin{gather}
U = U(x^-) \Rightarrow A^U_+ = U^{-1}A_+U + i U^{-1}\partial_+ U = 0.
\end{gather}
This symmetry enforces us to consider only a subsector of the whole Hilbert space, $\mathcal{H}_G \subset \mathcal{H}$, that contains only global singlets under the action of global $G$ group. In that Hilbert space we can introduce Hamiltonian and momentum operator as
\begin{gather}
 H = \int dx^+ \left[- \gtwo J^i \frac{1}{\partial_+^2} J^i  - \frac{i m^2_{\rm adj}}{4} \psi^i \frac{1}{\partial_+} \psi^i   \right], \quad P = \int dx^+ \left[\frac{i}{2}\psi^i_+ \partial_+ \psi^i_+ + i \bar{v}^\alpha\partial_+v^\alpha\right], \label{eq:HP}
 \end{gather}
the mass spectrum of the states could be find via relation $M^2= 2HP$. One can see that since $H$ and $P$ are singlet with respect to the gauge group $G$ they will correctly map $\mathcal{H}_G$ to itself and therefore their action is well established.  Now we will be only interested in the symmetries of the spectra of this Hamiltonian.
\section{Hamiltonian approach}
In this section we show explicitly how the Hilbert space of the model possesses a supersymmetric operator and generalize to the case when additional fermions are added to the model. We will mostly review the results of the paper.
During this section we will omit the subscripts $+$ in the notation of operators \eqref{eq:comrel} for the sake of brevity. The Hilbert space is constructed with the use of the following operators
\begin{gather}
\left\{\psi_i(p),\psi_j(q) \right\} = 2\pi\delta_{ij}\delta(p+q)
,\quad \left\{\bar{v}_\alpha(p),\bar{v}_\beta(q)\right\} = 2\pi\delta_{\alpha\beta}\delta(p-q),
\end{gather}
we will use a Schrödinger approach to this quantum problem --- the operators do not evolve with time while states do. Again, the algebra $\mathfrak{g}$ is represented by Hermitian matrices $T^i$ that satisfy the relation
$\left[T^i,T^j\right] = i f_{ijk} T^k$ and represented by matrices $\tau^i_{\alpha\beta}$ in the representation $\mathcal{R}$ of massless fermions $v_\alpha$. We define the following Hermitian current operators
	\begin{gather}
	J^\psi_i(p) = -\frac{i}{2} f_{ijk} \int \frac{dp_1}{2\pi} \psi_j(p+p_1) \psi_k(-p_1), \\
	J^v_i(p) = \int \frac{dp_1}{2\pi} \bar{v}_\alpha(p_1) \tau^i_{\alpha\beta} v_\beta(p_1+q), \label{eq:addcur}
	\end{gather}
	as usual when we deal with infinite numbers of operators involved we should take care how we define the operators in order to get meaningful results. For this case we define these operators using normal ordering.
	We notice immediately, that these operators are equal to the normal ordered versions of themselves --- there is no disambiguities in the definitions
	$:J_i^\psi(p): = J_i^\psi(p)$. It is easy to check that they give the right action in the Hilbert space. Namely, the commutator of currents with the field
	\begin{gather}
	[J^\psi_i(p),\psi_j(q)] = i f_{ijl} \psi_l(p+q),
	\end{gather}
	that shows that $\psi_j(q)$ transforms in the adjoint representation of the Lie algebra $\mathfrak{g}$ and $J^\psi_i(p)$ are indeed current operators of the Lie algebra and provide the correct representation of Kac-Moody algebra. Namely, let us check the commutators of two current operators
	\begin{gather*}
	[J^\psi_i(p), J^\psi_j(q)]=\frac{1}{2} f_{jmn} \int \frac{dp_1}{2\pi} \left(f_{ima}\psi_a(p+p_1+q)\psi_n(-p_1) + f_{ina}\psi_m(p_1+q)\psi_a(p-p_1) \right), \label{eq:curcom}
	\end{gather*}
	that contains some disambiguities related to normal ordering that must be resolved. Thus, let us assume for a while that $p+q \neq 0$, where we do not have any issues with normal ordering. Then can shift the integral in the second term and using Jacobi identity we get
	\begin{gather}
	[J^\psi_i(p), J^\psi_j(q)]
	=\frac12 \int \frac{dp_1}{2\pi} f_{anm}f_{ijm} \psi_a(p+p_1+q)\psi_n(-p_1) = i  f_{ijm} J^\psi_m(p+q).
	\end{gather}
	If $p=-q$ the commutator \eqref{eq:curcom} the integrand is ill-defined. Namely, by naive integration one can assume that these two terms coincide and therefore their difference should be zero. Actually, these two terms diverge and to take them into account correctly we should regularize the integral and after that perform the shifts. The other way to see this is to notice that the commutator
	\begin{gather}
	[J^\psi_i(p), J^\psi_j(-p)] = \notag\\
	=\frac{1}{2}\int \frac{dp_1}{2\pi} \left(f_{jmn}f_{ima} \psi_a(p_1) \psi_n(-p_1) + f_{jam}f_{imn}\psi_a(p_1-p) \psi_n(p-p_1)\right)
	\end{gather}
	is normal ordered (or both normal disordered) when $p_1 > p$ or $p_1 < 0$. But if $0<p_1 < p$ the first term is normal ordered while the second one is not. Because of that we are not allowed to perform a shift. By correctly performing the normal ordering we get some constant contribution proportional to $p$. Namely, after some algebra we finally at the following expression
	\begin{gather}
	[J^\psi_i(p),J^\psi_j(-q)] = i f_{ijm}J^\psi_m(p-q) +  \frac{C_f}{2} \delta_{ij} \delta(p-q),\quad f_{\alpha\beta i} f_{\alpha\beta j} = C_f \delta_{ij}
	\end{gather}
	Now at this level we can see that the Hilbert space spanned by the fermions in the adjoint representations already is supersymmetric. For this purpose we introduce the charge $Q^\psi$
	\begin{gather}
	Q^\psi = \frac13 \int \frac{dp_1}{2\pi} \psi_i(-p_1) J^\psi_i(p_1),
	\end{gather}
	one can check that this operator is normal ordered $: Q^\psi : =  Q^\psi$. Anticommutator with the fields $\psi_i(p)$ reads as
	\begin{gather}
 	\left\{Q^\psi,\psi_i(p)\right\}
 	=\frac13 \int \frac{dp_1}{2\pi}\left[i \psi_j(-p_1) f_{jik}\psi_k(p+p_1) \right] + \frac13 J^\psi_i(p) = J^\psi_i(p),
	\end{gather}
	and since we do not shift integrals and there are not disambiguities and we can trust this result. Now consider commutator with $J^\psi_i(p)$. We have
	\begin{gather*}
	[Q^\psi, J^\psi_i(p)]
	= -\frac{p C_f}{12\pi} \psi_i(p) + \frac{i}{3} \int \frac{dp_1}{2\pi} \left(f_{jik}\psi_k(p-p_1) J^\psi_j(p_1) + \psi_j(-p_1) f_{jik}J^\psi_k(p+p_1) \right),
	\end{gather*}
	Again to reconcile them we should make a shift. As in the case of co-cycle for the Kac-Moody algebra we should perform it in a very cautious way.  We notice that this terms again both either ordered or disordered when $p_1<-p$ or $p_1>0$, then we get
	\begin{gather}
	[Q^\psi,J^\psi_i(p)] = - \frac{p C_f}{4\pi}  \psi_i(p),
	\end{gather}
	the action of $Q^\psi$ coincide with the one derived by Kutasov et al. We generalize this construction when additional fermions are involved. To make the computation quite general we assume that we added just some additional system of currents $J^v_i$ with the following commutation relations
	\begin{gather}
	[J_i^v(p), J_j^v(q)] =i f_{ijm} J^v_m(p+q) + \frac{k}{2} p\delta(p+q) \delta_{ij},
 	\end{gather}
 	that could come from some other reasons not necessarily by introduction of the massless fermions.
 	We define the total current $J_i = J^\psi_i + J^v_i$ and since $J^\psi$ and $J^v$ commute we immediately get
 	\begin{gather}
 	[J_i(p), J_j(q)] =i f_{ijm} J_m(p+q) + \frac{k_t}{2} p\delta(p+q) \delta_{ij},\quad k_t = C_f + k.
 	\end{gather}
 	Then we introduce new supersymmetric charge as
 	\begin{gather}
 	Q = \int \frac{dp_1}{2\pi} \psi_i(-p_1) \left[ \frac13 J^\psi_i(p_1) + J^v_i(p_1)\right], \label{eq:Qoper}
 	\end{gather}
 	note that a similar expression was suggested in \cite{kazama1989characterization,kazama1989new} . It is easy to check that the same commutation relations hold for the new charge $Q$
 	\begin{gather}
 	\left\{Q,\psi_i(q)\right\} = J_i(q) = J^\psi_i(q)+J^v_i(q)
 	\end{gather}
 	After that we try to commute it with the current $J^t_i(q)$, we have
 	\begin{gather}
 	\left[Q, J^\psi_i(q) + J^v_i(q)\right]  = \left[Q^\psi,J^\psi_i(q)\right] + \int \frac{dp_1}{2\pi}\left[\psi_j(-p_1) J_j^v(p_1),J^\psi_i(q)+J^v_i(q)\right],
 	\end{gather}
 	the first term was computed before. The second term could be computed in the similar fashion and we arrive at
 	\begin{gather}
 	\left[Q,J_i(q)\right] = - \frac{k_t q}{4\pi} \psi_i(q).
 	\end{gather}
 	Now we can easily check that the $Q$ commutes with the Hamiltonian for a particular mass $m$ and momentum operator defined in the previous section. The $[Q,P]=0$ just because $Q$ does not carry any momentum or just noticing that it does not depend on the coordinates $x^-$ explicitly. The only thing is left to check that $[Q,H]=0$. We notice that Hamiltonian in the momentum representation has the following form
 	\begin{gather}
 	H = \int \frac{dp}{2\pi} \left[ g_{\rm YM}^2 \frac{J^i(p) J^i(-p)}{p^2} + \frac{m^2_{\rm adj} \psi^i(p) \psi^i(-p)}{4p}\right]
 	\end{gather}
 	Then commuting it with $Q$ we are getting
 	\begin{gather}
 	\left[Q,H\right] = \int \frac{dp}{2\pi} \left[- k_t \frac{g_{\rm YM}^2}{2\pi} \frac{\psi^i(p) J^i(-p)}{p} + \frac{m^2_{\rm adj} \psi^i(p) J^i(-p)}{2p}\right],
 	\end{gather}
 	and for $m^2 = \frac{k_t g^2_{\rm YM}}{\pi}$ we would get that $Q$ is a fermionic symmetry of our system. For instance if we consider $G=SU(N)$ and one adjoint fermion, then $k=N$ and the mass is $m^2 = \frac{\gtwo N}{\pi}$. If we add $N_f$ fermions in the fundamental representation to the Hamiltonian the mass of the adjoint fermion should be
 	\begin{gather}
 	m^2_{\rm adj} = \frac{g_{\rm YM}^2 (N+N_f)}{\pi},
 	\end{gather}
 	to make the whole system respect the fermionic symmetry $Q$. To state that the fermionic operator $Q$ is a real supersymmetric operator we should compute $Q^2$ and compare with $P$, we will do this computation in the next section. It would be quite interesting to check this result numerically. For a while, we can check this claim numerically only in the large $N$ limit. If we fix $N_f$, the correction would be small in the large $N$ limit. Therefore we wouldn't be able to see directly small $N_f$ correction to the mass of adjoint fermion. Nonetheless we can check that the leading correction is left unaffected and supersymmetry indeed arises in the large $N$ limit at the same mass\footnote{I would like to thank Ross Dempsey for providing numerical data, that confirms this claim.}. The other approach would involve to consider large number of massless fermions in the fundamental representation $N_f \sim N$, that would be quite hard to implement. And the final consideration would involve just to add a massless fermion in an adjoint representation, that should double the value of the supersymmetric mass. Also, this result is quite similar to the universality of QCD models with massless fermions proposed by Kutasov and Schwimmer \cite{kutasov1995universality}, where the spectrum of massive states does not depend on the concrete structure of the massless sector but only on the coefficient in front of WZW action.

 	One might wonder how would additional massless fermion drastically change the properties of the matter. More surprisingly, why does the spectra of the model does not depend on the actual structure of the massless sector of the theory, but only on one colloquial factor $k$, that determines the level of Kac-Moody algebra. The second question was partially addressed in the case when all fermions are massless \cite{kutasov1995universality}. To see how both of these question could be answered, let us consider the case of $U(1)$ gauge theory, that was partially reviewed in the introduction. Namely, we want to consider an electromagnetic field coupled to a $N_f$ massless fermions and one massive fermion that is not coupled to the electromagnetic field. In this case we have the following action
 	\begin{gather}
 	S = \int d^2 x \left[-\frac{1}{4g^2} F_{\mu\nu}^2 + i \bar{\psi}\gamma^\mu \partial_\mu \psi + i m \bar{\psi}\psi+ \sum_j i\bar{\chi}_j\gamma^\mu\left(\partial_\mu + i q_j A_\mu\right)\chi_j \right],
 	\end{gather}
 	any gauge field could be decomposed as $A_\mu = \partial_\mu \alpha + \epsilon_{\mu\nu} \partial_\nu \beta$. The pure gauge term could be removed by a usual gauge rotation, while $\beta$ term could be removed via the axial rotation $\chi_j \to e^{iq_j\beta \gamma^5} \chi_j$. The last rotation is not respected by a fermion measure. We have to add Schwinger term to the action. After that we arrive at the following action
 	\begin{gather}
 	S = \int d^2 x \left[\frac{1}{2 g^2} \left(\Delta \beta\right)^2-k \beta \Delta \beta +  i \bar{\psi}\gamma^\mu \partial_\mu \psi + i m_f \bar{\psi}\psi+ \sum_j i\bar{\chi}_j\gamma^\mu\partial_\mu \chi_j  \right],
 	\end{gather}
 	after that we see that the field $\beta$ has two poles: one with mass $k^2 = g^2\sum_j q_j^2$ with positive residue and $k^2=0$ with negative residue. The first one corresponds to a real bosonic propagating degree of freedom, the other one is non-physical.  Its role is the cancellation of one of the massless degrees of freedom and instead of $N_f$ massless fermionic degrees of freedom we have $N_f-1$ massless fermions. To be completely right, one can bosonize these $N_f$ fermion and get $SU_1(N_f) \times U_{N_f}(1)$ WZW model. The factor $U_{N_f}(1)$ is coupled to the gauge field and essentially creates mass for a photon $m^2=g^2 k$ . Thus we get $SU_1(N_f)$ WZW together with a massive photon. After that if we pick $m_f^2 =  g^2 k$ we would have that fermionic and bosonic degrees of freedom will have the same mass and therefore we will get trivially a supersymmetric spectrum. 

 	One important lessons we should draw, that while the spectrum of the massive states is supersymmetric (for each massive bosonic state we have a massive fermionic state), while the spectrum of massless states is not.  We expect the same to hold for non-abelian case.  We need an additional investigation to figure out the concrete structure of the massless states \cite{Delmastro:2021otj}. Nevertheless, it could be possible that the massless part of the spectra is still supersymmetric. Thus, in the case of one massless fermion in fundamental representation and one massive fermion in the adjoint representation, it is known that there are massless baryon states in the spectrum, in addition to a massless meson. If $N$ is odd than the Baryon number 1 state is a fermion and could be a partner of the meson \cite{KlebanovPrivate}. So the spectrum would be completely supersymmetric.

 	\section{Path Integral Derivation}
 	The previous approach for the abelian case could be generalized to the non-abelian case. 
 	In this section we rederive the results of the previous section by using path integral approach. It has an advantage, because would allow to consider and find these supersymmetric transformations not only in the light-cone quantization.  As it was discussed in the introduction to make a WZW model supersymmetric we should just simply add a massless fermion in the adjoint representation. This statement could be formulated at the level of path integral and explicitly write down the transformation rules for the WZW field $g$, that is an element of a gauge group, and fermionic field $\psi^i$. Namely, if we have a WZW model at level $k$ and massless adjoint Majorana fermion \cite{Witten:1991mk}:
	\begin{gather}
	S_0 = \int d^2 x \tr\left[ i \psi_+ \partial_- \psi_+ +  i \psi_-\partial_+ \psi_- \right] + k W[g], \label{eq:actred}
	\end{gather}
	then one can check that the action \eqref{eq:actred} possesses the following symmetry
	\begin{gather}
	\delta g = \frac{4\pi i}{k}\left[ \epsilon_- g \psi_+ +  \epsilon_+ \psi_- g\right],\notag\\
	\delta \psi_+ = \epsilon_-\left( g^{-1}\partial_+ g - \frac{4 \pi i}{k} \psi_+^2\right),\quad \delta \psi_- = \epsilon_+ \left(\partial_- g g^{-1} +  \frac{4 \pi i}{k}\psi_-^2\right). 	\label{eq:trans}
	\end{gather}
	Indeed, let us check for $\eps_+=0$ that the action is  invariant
	\begin{align}
	&\delta S_f =2 i \epsilon_- \int d^2 x \Tr\left[ g^{-1} \partial_+ g\partial_- \psi_+\right] \label{deltaf}\\
	&\delta S_{\rm WZW}(g) =  \frac{k}{2\pi} \int d^2 x \Tr\left[g^{-1} \delta g \partial_- \left(g^{-1} \partial_+ g\right)\right] 
	=2i\eps_- \int d^2x \tr\left[\psi_+ \partial_-\left(g^{-1}\partial_+ g\right)\right] \label{deltawzw}
\end{align}
Combining these two variations \eqref{deltawzw} and \eqref{deltaf} we would get a total derivative. One important observation in this derivation is that the transformation \eqref{eq:trans} works for any chosen $k$. It is deeply connected to the fact found in the previous section: that for 2D QCD with massive adjoint fermions and a massless fermions we can always fine-tune mass  $m_{\rm adj}$ that the model becomes supersymmetric.

Now we will show that the symmetry \eqref{eq:trans} is responsible for the supersymmetric transformations. To do this, we would like to make the following simple transformation of the action
\eqref{eq:2dqcdaction}, that would make it look similar to the $\mathcal{N}=1$ supersymmetric WZW. Thus, we pick gauge $A_- =0$ as in the previous case, but notice that we can always find an element of the group $g \in G$ such that
\begin{gather}
A_-=0,\quad A_+ = -i g^{-1}\partial_+ g,\quad \D_-=\partial_-,\quad \D_+ = g^{-1}\cdot \partial_+ \cdot g,
\end{gather}
that allows to rewrite the action in the following way
\begin{gather}
S = \int d^2x \tr \left[-\frac{1}{2 g_{\rm YM}^2} \left(\partial_- \left( g^{-1}\partial_+ g\right)\right)^2+ i \psi_+ \partial_- \psi_+ + i g \psi_- g^{-1} \partial_-\left(g \psi_- g^{-1}\right)  + i \sqrt{2} m_{\rm adj} \psi_+ \psi_-+\right.\notag\\\left.+ i \bar{v}_+ \partial_- v_+ + i \bar{v}_- \rho(g^{-1}) \partial_+ \left(\rho(g)v_- \right) \right],
\end{gather}
we can easily get rid of $g$ in the action by making rotations $v_- \to \rho(g)v_-$ and $\psi_- \to g^{-1}\psi_- g$, that could be done without big troubles, but the price we should pay is to add to the action the WZW action, because the measure of fermions is not invariant under this transformations. After such a rotation, the massless fermions "decouple" from the action. These massless states are still present, but they are completely decoupled from the interaction with the gauge field, while the interacting degrees of freedom could be represented with the use of the WZW action. For instance, if we consider $G=SU(N)$ and consider $N_f$ fundamentals there is a well-known \cite{gogolin2004bosonization}
\begin{gather}
S = \sum^{N_f}_{\alpha=1} \int d^2x \bar{\psi}_{i\alpha} \slashed \partial \psi_{i\alpha} \leftrightarrow \int d^2 x \left[ N_f W[g]+N_c W[h] + \left(\partial \phi\right)^2 \right],
\end{gather}
where $g \in SU(N), h\in SU(N_f)$ and $\phi$ representes overall phase. One can see that the gauge fields interact only with $W[g]$ while $W[h]$ and $\phi$ decouple from the interaction.

Let us note that in the axial gauge, the F.Popov ghosts are decoupled from the rest of the system and thus could be ignored.  So at the end we arrive at the following action
\begin{gather}
S = \int d^2 x \tr\left[ i \psi_+ \partial_- \psi_+ +  i \psi_-\partial_+ \psi_- \right] + k W[g] + \notag\\
+\int d^2 x \tr\left[i \sqrt{2} m_{\rm adj} \psi_- g \psi_+ g^{-1} - \frac{1}{2 g_{\rm YM}^2}\left(\partial_- \left[g^{-1} \partial_+ g\right]\right)^2  \right],\label{eq:fullaction}
\end{gather}
where coefficient $k=c_f+k_0$ in front of the WZW action comes separately from the adjoint fermions $c_f$ and massless part $k_0$. The action is very similar to the one considered above and therefore it is natural to conjecture that the same transformation \eqref{eq:trans} leaves the action \eqref{eq:fullaction} invariant. The direct computation shows that we should add the following terms to the transformation to make everything consistent (for brevity we consider only $\eps_-=0$, the other part of transformations could be easily written in a similar fashion)
\begin{gather}
\delta g = \frac{4\pi i}{k} \eps_- g \psi_+,\quad
\delta\psi_- = \eps_- \frac{\pi m_{\rm adj}}{\sqrt{2}g^2 k}g F_{+-} g^{-1} \notag\\
\delta \psi_+ = \eps_- \left(g^{-1} \partial_+ g - \frac{4 \pi i}{k} \psi_+^2\right) + \frac{\pi}{g_{\rm YM}^2 k}\eps_- \left(\frac{1}{2}m^2_{\rm adj} g^{-1} \partial_+ g -  \mathcal{D}_+ F_{-+}\right),
\end{gather}
we can easily check that under this transformations the whole action is invariant. Indeed, the variation of the Yang-Mills part of the action is
\begin{align}
&\delta\left(g^{-1} \partial_+ g\right) = \frac{2\pi i}{k} \eps_- \left[\partial_+ +  \ad\left( g^{-1}\partial_+ g\right)\right]\psi_+, \quad \ad(g^{-1}\partial_+ g)\psi_+ = \left[g^{-1}\partial_+ g,\psi_+ \right]  \notag\\
&\delta S_{YM} =
\frac{4 \pi i\epsilon_- }{g_{\rm YM}^2 k}\itd \Tr \left[\mathcal{D}_+ F_{-+} \partial_-\psi_+\right],
\end{align}
Then we have the following additional terms in the fermion part of the action
\begin{align}
&\delta S_- = \int d^2 x \eps_- \Tr\left[\frac{2\pi i m_{\rm adj}}{\sqrt{2}g_{\rm YM}^2 k}g F_{+-} g^{-1} \left[2\partial_+ \psi_- +  \sqrt{2}m_{\rm adj} g\psi_+ g^{-1}\right]\right],\notag\\
&\delta S_+ =  \frac{2\pi i}{g_{\rm YM}^2 k} \int d^2 x \eps_- \Tr\left[\left(\frac12 m^2_{\rm adj} g^{-1} \partial_+ g -  \mathcal{D}_+ F_{-+}\right) \left[  2\partial_- \psi_+ - \sqrt{2}m_{\rm adj} g^{-1} \psi_- g\right]\right],\notag\\
&\delta S_m 
= - \sqrt{2} i \eps_- m_{\rm adj} \int d^2 x \Tr\left[\psi_- \partial_+ g g^{-1}\right] \label{eq:allvar}
\end{align}
Combining all $\delta S_-, \delta S_+, \delta S_{YM}$ we arrive at the following variation
\begin{gather}
\delta S = \frac{\sqrt{2}\pi i m^3_{\rm adj}\epsilon_- }{g_{\rm YM}^2 k}\Tr \left[ \psi_- \partial_+ g g^{-1}\right],
\end{gather}
where we have integrated by parts and used that $\mathcal{D}_+\left(g^{-1} \psi_- g\right) =g^{-1} \partial_+ \psi_- g$.
That cancels out by $\delta S_m$ \eqref{eq:allvar} if the following condition is satisfied
\begin{gather}
\frac{\sqrt{2}\pi m^3_{\rm adj}}{g_{\rm YM}^2 k} = \sqrt{2}m_{\rm adj},\quad m^2_{\rm adj} = \frac{g_{YM}^2 k}{\pi} \label{masssum},
\end{gather}
that coincides with the results of the previous section. Now let us compute the square of the $Q$ operator \eqref{eq:Qoper} 
\begin{gather}
 	\left\{Q,Q\right\} = \frac{k + c_f}{4\pi} \int \frac{dp_1}{2\pi}\psi_i(-p_1) p_1 \psi_i(p_1) + \int \frac{dp_1}{2\pi}J_i^v(-p_1) J_i^v(p_1),
 \end{gather}
 while it is easy to see that it does not coincide with the momentum operator of the whole system, nonetheless it reproduces the momentum operator of the interacting subpart described by the action \eqref{eq:fullaction} \footnote{I would like to thank S.Pufu and I.Klebanov for discussion and checking this relation}. It shows that $Q$ is indeed the supersymmetry of QCD with adjoint fermions and deeply connected to the $\mathcal{N}=1$ supersymmetric WZW models. One can notice that the supersymmetric transformation considered in this section looks like a gauge transformation with gauge parameter $\eps_- \psi_+$ (in two dimensions the supersymmetric transformation of gauge theories could indeed be casted in such a form). The reason why in this case the supersymmetry demands the additional mass term for the adjoint fermions is a sensitivity to the chiral transformations of the measure. The original supersymmetry transformations breaks the chiral symmetry and to take into account the change in the measure we should add to the action the additional terms (that is just a variation of the WZW term), that should and could be compensated by the mass term for fermionic fields.

\section{Discussion and possible generalizations}

One of the interesting generalizations of the proposed mechanism of supersymmetric gauge theories in two dimensions, would involve the use of the coset construction of the $\mathcal{N}=1$ supersymmetric WZW models \cite{Witten:1991mk,kazama1989characterization,kazama1989new}. It is very well-known that such constructions could lead to the $\mathcal{N}=2$ supersymmetry. The supersymmetric action in this case has the following form
\begin{gather}
 S = k W[g] + \frac{1}{2\pi} \int d^2 z \tr \left[B_- g^{-1}\partial_+ g - B_+ \partial_- g g^{-1} - B_- B_+ + B_- g^{-1} B_+ g \right] + \notag\\
 \frac{i}{4\pi} \int d^2 z \tr \left[\psi_+ \mathcal{D}^B_- \psi_+ + \psi_- \mathcal{D}^B_+ \psi_+ \right],
\end{gather}
where $B$ is a gauge field that values in the subalebgra $\mathfrak{h} \subset \mathfrak{g}$ and $\psi_\pm$ belong to the subspace $\mathfrak{g}/\mathfrak{h}$. Under general assumptions this model would become $\mathcal{N}=2$ supersymmetric. It would be very interesting to use this supersymmetry to construct $\mathcal{N}=2$ QCD model. Nevertheless, the obstacle includes the introduction of the additional gauge field $B_\pm$ that would also gauge the original gauge field.
\section{Acknowledgement}

We thank Igor Klebanov for the proposed problem and valuable discussion and S.Dubovsky, J.Sonnenschein, D.Kutasov, Y.Wang and S.Pufu for very illuminating questions and discussions. The author is grateful to O.Diatlyk and C.Jepsen for useful discussions and comments on the drafts.
F.K.P. is currently a Simons Junior Fellow at NYU and supported by a grant 855325FP from the Simons Foundation. We are thankful to KITP for hospitality and all participants of the program "Confinement, Flux Tubes, and Large N", where the final stages of the project were completed.

\appendix

\section{Notations and conventions}
In this section we put all conventions and notations that have been used throughout the main body. In two dimensional space-time it would be convinient to study it in the light-cone coordinates
\begin{gather}
ds^2 = 2 dx^+ dx^-,\quad x^\pm = \frac{t \pm x}{\sqrt{2}}, \quad \eta_{+-}=\eta_{-+}=1,\quad   \gamma^+=\left(\gamma^-\right)^T=\begin{pmatrix}
0 & \sqrt{2} \\
0 & 0
\end{pmatrix}.
\end{gather}
We represent our gauge group $G$ with the use of its Lie algebra $\mathfrak{g}$, assuming that it is finite-dimensional, and has the following basis of the Hermitian matrices $T^i$:
\begin{gather}
T^i \in \mathfrak{g},\quad \left[T^i,T^j\right] = i f_{ijk}T^k, \quad \tr\left[T^i T^j\right] = \frac12\delta^{ij},
\end{gather}
the structure constants $f_{ijk}$ are completely antisymmetric and satisfy the Jacobi identity.

The fermion field in the adjoint representation could be decomposed with respect to this basis in the following way
\begin{gather}
	\Psi=\Psi^i T^i,\quad \Psi^i = \frac{1}{2^\frac14}\begin{pmatrix}
	\psi^i_+ \\
	\psi^i_-
	\end{pmatrix} \in \mathbb{R}^2, \quad \bar{\Psi}=\Psi^T \gamma^0.
\end{gather}
The massless fermion field is assumed to be in some representation (that is not necessarily should be an irreducible representation)
\begin{gather}
	v = \begin{pmatrix}
	v_+ \\
	v_-
	\end{pmatrix}, \quad v_\pm = \left\{ v_\pm^\alpha \right\} \in \mathbb{C}^n=\mathcal{R},\quad \bar{v} = v^\dagger \gamma^0, \notag\\
	\rho_\mathcal{R} : \mathfrak{g} \to \operatorname{End}\left(\mathcal{R}\right),\quad \rho_\mathcal{R}(T^i)^\alpha_\beta = \tau^{i \alpha}_{\beta}
\end{gather}
The gauge field $A_\mu \in \mathfrak{g}$ belongs to the adjoint representation of the group $G$ and again could be decomposed with the use of the basis $T^i$
\begin{gather}
 A_\mu = A_\mu^i T^i,
\end{gather}
the covariant derivative is defined in the following way
\begin{alignat}{2}
&\D_\mu=\partial_\mu + i A_\mu, &&F_{\mu\nu}=-i\left[\D_\mu,\D_\nu\right] = \partial_\mu A_\nu - \partial_\mu A_\nu + i [A_\mu,A_\nu] = F^i_{\mu\nu} T^i, \notag\\
&\D_\mu \Psi = \partial_\mu \Psi + i\left[A_\mu,\Psi\right],&& \D_\mu v = \partial_\mu  v+ i \rho(A_\mu) v, \notag\\
&\left(\D_\mu \Psi\right)^i = \partial \Psi^i - f_{ijk}A^j \Psi^k, &&\left(D_\mu v\right)^\alpha=\partial_\mu v^\alpha + i A_\mu^i \tau^{i\alpha}_{\beta}v^\beta, \notag\\
&F^i_{+-} =  \partial_+ A^i_- - \partial_- A^i_+ - f^{ijk} A^j_+ A^k_-,&& \notag
\end{alignat}
where $F^i_{+-}$ is the only non-zero component of the curvature $F_{\mu\nu}$ in 2 dimensions.

\bibliographystyle{ssg}
\bibliography{biblio}
\end{document}